\documentclass[twocolumn,aps,showpacs,superscriptaddress,showkeys,floatfix]{revtex4} 
\usepackage{graphicx}
\usepackage{color} 
\begin{document}
\title{
Numerical Modeling of Radiation-Dominated and
  QED-Strong Regimes of Laser-Plasma Interaction } 
\author{Igor V. Sokolov}
\email{igorsok@umich.edu}
\affiliation{Space Physics Research Laboratory, University of Michigan, Ann
Arbor, Michigan 48109, USA }
\author{Natalia M. Naumova}
\affiliation{Laboratoire d'Optique Appliqu\'{e}e, 
UMR 7639 ENSTA, Ecole Polytechnique, CNRS, 91761 Palaiseau, France}
\author{John A. Nees}
\affiliation{Center for Ultrafast Optical Science and FOCUS Center, University
of Michigan, Ann Arbor, Michigan 48109, USA}
\date{\today}
\begin{abstract}
Ultra-strong laser pulses can be so intense that an
electron in the focused beam loses significant 
energy due
to $\gamma$-photon emission while its motion deviates
via the radiation back-reaction. 
Numerical methods and tools designed to simulate radiation-dominated and 
QED-strong laser-plasma interactions are summarized here.
\end{abstract}


\maketitle
\section{Introduction}
Progress in laser technologies has resulted in the opportunity 
to create ultra-strong electromagnetic fields in tightly focused laser beams.
In the present paper we discuss the numerical methods
designed to simulate
processes in strong
pulsed laser fields 
interacting with
plasma. 
Attention is
paid to the recently achieved range of intensities, $J\ge
2\cdot10^{22}{\rm W/cm}^2$  
\cite{1022}, and the larger intensities projected, $J\sim
10^{25}{\rm W/cm}^2$ \cite{ELI}.
 
For a typical laser 
wavelength, $\lambda\sim1\mu$m, electron motion in laser fields at
$J\ge 10^{18}{\rm W/cm}^2$ 
is relativistic:
\begin{equation}\label{eq:strong}
|{\bf a}|\gg1,\qquad{\bf a}=\frac{e{\bf A}}{m_ec^2},
\end{equation} 
where 
$m_e$ and $e=-|e|$ are the mass and the electric charge of electron.

However, if we want to evaluate the properties of an electron in a strong
field as an {\it emitting} particle, moreover, a particle, which emits
{\it photons} we need to be guided by the more severe condition:
\begin{equation}\label{aalpha}
|{\bf a}|_\alpha=\alpha|{\bf a}|\ge 1,
\end{equation}
in which the fundamental fine structure constant is present, $\alpha =
e^2/(c\hbar)\approx1/137$, linking its radiation to its motion
(herewith, the subscript $\alpha$ denotes the dimensionless
parameter multiplied by $\alpha$). With 
the recently achieved intensity of $J\sim
2\cdot10^{22}\ {\rm W/cm}^2\sim (1/\alpha^2)\cdot10^{18}\ {\rm
  W/cm^2}$, this newly important dimensionless
parameter exceeded unity! 

However, this estimate could be applicable only to a 'theoretical
laser', for which the photon energy, $\hbar\omega$, would be comparable 
to the 'characteristic' atomic unit
of energy, ${\rm 2Ry}=\alpha^2m_ec^2$. For a real laser,
in addition to the field magnitude, 
importance rests on 
the laser
photon energy normalized 
by ${\rm 2Ry}$:
\begin{equation}
\omega_{\rm 2Ry}=\frac{\hbar\omega}{\rm 2Ry}\approx\frac{45 {\rm
    nm}}\lambda,\qquad {\rm 2Ry}=\alpha^2m_ec^2\approx 27.3{\rm eV}.
\end{equation}
For the majority of ultra-strong lasers this parameter is of the order of
$10^{-1}$: $\omega_{\rm 2Ry}\sim0.04$ for the 
Nd-glass laser ($\lambda\approx1.06\ \mu$m), $\omega_{\rm 2Ry}\sim0.06$ for
the Ti-sapphire laser ($\lambda\approx0.8\ \mu$m).  
Therefore, the following product,
\begin{equation}\label{aomega}
\omega_{\rm 2Ry}
\left|\frac{d{\bf a}}{d\xi}\right|_\alpha=\sqrt{\frac{J}{J_p}},
\quad J_p=\frac{cE_p^2}{4\pi}\approx
2.4\cdot
    10^{25}{\rm W/cm}^2,
\end{equation}
is less than one even for planned intensities (although $|d{\bf
  a}/d\xi|_\alpha$ might be greater than one).  
Herewith, the estimates are made for the 1D wave field, 
${\bf a}={\bf a}(\xi)$, $\xi=\omega (t-x/c)$,  $0\le\xi\le\xi_{\rm
max}$. Eq.(\ref{aomega}) is expressed in terms of
the local instantaneous intensity of the laser
wave, $J$. Note that the Left Hand Side (LHS) of Eq.(\ref{aomega})
equals the ratio, $|{\bf E}|/E_p$, of the wave electric field, $|{\bf
  E}|=\sqrt{4\pi J/c}$, to the
characteristic field, $E_p=|e|/\lambdabar_C^2$, constructed from an
elementary charge and the Compton
wavelength:
$$ 
\lambdabar_C=\frac\hbar{m_ec}\approx 3.9\cdot10^{-11}{\rm cm},\quad
\lambda_C=2\pi\lambdabar_C\approx 2.4\cdot10^{-10}{\rm
cm}.
$$
This field strength is associated with the Coulomb field between
the components of a virtual electron-positron pair (which are
'separated' by the Compton wavelength). 
Across the interval of intensities bounded by Ineq.(\ref{aalpha}) from
below and by Eq.(\ref{aomega}) from above, that is,
 at
\begin{equation}\label{interval}
\omega_{\rm 2Ry}\approx\frac{45 {\rm
    nm}}\lambda\le\sqrt{\frac{J}{J_p}}\le1
\end{equation}
the role of 
important 
physical effects changes dramatically, incorporating radiation and its
back-reaction, and QED effects of electron recoil and spin as well as 
pair production. Given that currently available laser intensities can
access this kind of interaction, it is clear that the development of a
suitable model is timely.

{\bf 
Radiation-dominated laser fields.} An accelerated electron in a strong laser field emits 
high-frequency radiation. The radiation back-reaction decelerates such an
electron, the effect being more pronounced for longer laser pulses
\cite{our}. In  Ref.\cite{kogaetal} 
a condition for the field to be {\bf radiation-dominated} is formulated in terms of the ratio between the magnitudes of the
Lorentz force and of the radiation force, which gives: 
\begin{equation}\label{raddom}
\frac23\omega_{\rm 2Ry}({\cal E}-p_\|)_\alpha\left(\left|\frac{d{\bf a}}{d\xi}\right|_\alpha\right)^2\ge1.
\end{equation} 
Herewith the electron dimensionless energy, 
${\cal E}$, and
its momentum, ${\bf p}$, are related to $m_ec^2$, and $m_ec$ correspondingly,
and subscript $\|$ herewith denotes the vector projection on the direction of the
wave propagation. 

While a 
strong laser pulse interacts with  energetic electrons, which 
move {\it oppositely} to the direction of the pulse propagation, the condition
${\cal E}-p_\|\approx 2{\cal E}\gg 1$,
facilitates the fulfillment of Ineq.(\ref{raddom}).  
In the course of a strong laser pulse interacting 
with a dense plasma the 
counterpropagating electrons may be accelerated in a backward direction by a 
charge separation field. For this reason, the radiation effects in the course 
of laser-plasma interaction are widely investigated (see
Refs.\cite{kogaetal,lau03,FI})  and efficient computational tools are in demand.
 
{\bf  
QED-strong laser fields.}
In Quantum Electro-Dynamics (QED) an electric field should be treated as  
strong if it exceeds the Schwinger limit: $|{\bf E}|\ge
E_S=m_ec^2/(|e|\lambdabar_C)$ 
(see Ref.\cite{schw}). Such field is potentially capable of separating a virtual 
electron-positron pair providing an energy, which exceeds the electron rest 
mass energy, $m_ec^2$, to a charge, $e=\mp|e|$, over an acceleration length as small 
as the Compton wavelength. 
Typical effects in QED-strong fields are high-energy photon 
emission from electrons or positrons and 
electron-positron pair creation from high-energy photons (see Refs.\cite{Mark,kb,ourPRL}). 

Here we assume that the field invariants (see \cite{ll}) are small as compared to the Schwinger field:
\begin{equation}\label{eq:smallinv}
|E^2-B^2|\ll E_S^2,\qquad |({\bf E}\cdot{\bf B})|\ll E_S^2,
\end{equation}
${\bf B}$ being the magnetic field. 
Below, the term ``QED-strong field'' is only applied to the field experienced by a particle. 
For example, a particle in the 1D
wave field, may experience a QED-strong
field,  
$E_0=|d{\bf A}/d\xi|\omega({\cal E}-p_\|)/c$, because the laser
frequency is Doppler
upshifted in the comoving frame of reference. 
The Lorentz-transformed field exceeds the Schwinger limit, 
if 
\begin{equation}\label{QED-strong}
\chi=\frac23 E_0/E_S
=\frac{2\lambdabar_C}{3\lambdabar}({\cal
 E}-p_\|)\left|\frac{d{\bf a}}{d\xi}\right|\gg1,
\end{equation}
where $\lambdabar= c/\omega$. 
Numerically, the parameter, $\chi$, equals:
$$
\chi=\frac32 \omega_{\rm 2Ry}({\cal
  E}-p_\|)_\alpha\left|\frac{d{\bf a}}{d\xi}\right|_\alpha
\approx 0.7\frac{({\cal
  E}-p_\|)}{10^3}\sqrt{\frac{J}{10^{23}[{\rm W/cm}^2]}}.
$$ 

{\bf Estimates for laser-driven electrons.} In the critical parameters
as in Ineqs.(\ref{raddom},\ref{QED-strong}) the factor, ${\cal E}-p_\|$ 
 is not linked to the wave intensity in the case
 where 
 electrons
are accelerated by 
an external source.  
In the course of the laser-plasma interaction, however, for bulk
electrons $({\cal E}-p_\|)\sim{\cal E}\sim |{\bf
  p}_\perp|$. As long as the radiation back-reaction does not dominate,
the conservation law for the generalized momentum of electron
gives: ${\bf p}_\perp \approx
  -{\bf a}$, and the LHS of
  Eqs.(\ref{raddom}) may be evaluated in terms of the vector
  potential amplitude, $a_0$. The wave becomes
  radiation-dominated (see Ineq.\ref{raddom}), if:
$$
a_0\ge(\omega_{\rm 2Ry})^{-1/3},
\quad 
J\ge J_p (\omega_{\rm 2Ry})^{4/3}\sim(3-5)\cdot 10^{23}{\rm
  W/cm}^2.
$$    
Less certain is the estimate for 
the 
significance of 
 QED effects. On one hand, for 
 fields just approaching the
radiation-dominated regime
 QED effects  
 are already not fairly 
 neglected:
$\chi\sim(3/2)(\omega_{\rm 2Ry})^{1/3}\approx(0.5-0.6)$. On the other hand, in
radiation-dominated fields the estimate for ${\cal E}$ 
that
we used above
is not reliable. 
Because of this complexity, we
surmise that the significance of 
QED effects
in this regime can only be verified by direct
numerical simulations.

{\bf 
Paper content and structure.}
Numerical simulations of  
laser-plasma interactions become increasingly complicated while 
proceeding to higher intensities. At intensities $J\ge2\cdot 10^{22}{\rm W/cm}^2$ the
model should incorporate the radiation back-reaction on the
emitting electron. 
In this
range 
$\chi\ll 1$ for bulk electrons, making the QED-effects
negligible. This model is presented in Sec. II. 
At 
$J\ge 3\cdot 10^{23}{\rm W/cm}^2$ the QED corrrections should be incorporated 
to achieve a quantitative accuracy for
electrons with $\chi\sim 1$. These corrections may be found in Sec.
III.  
At larger 
intensities, $J\ge10^{24}{\rm W/cm}^2$ the high-energy photons emitted
by electrons and positrons produce a macroscopically large number of
electron-positron pairs, as shown in Sec. IV.

In each section we first summarize the theoretical model. Then we
provide the analytical
solutions, which may be used to benchmark numerical models. After this we describe the elements of the
numerical scheme.
\section{QED-weak fields ($\chi\ll 1$)}
\subsection{Theoretical notes}
 \subsubsection{Emission spectrum}
In Ref.\cite{jack}
the spectral and angular distribution, 
$d{\cal E}_{\rm rad}/(d\omega^\prime d{\bf n})$,
of the radiation energy, emitted by 
an electron and related to the interval of
frequency, $d\omega^\prime$, and to the element of solid angle, $d{\bf
  n}$, for a  
polarization vector, ${\bf l}$, is described with the following formula:
\begin{equation}\label{eq:Jackson}
\frac{d{\cal E}_{\rm rad}(\omega^\prime,{\bf
  n},{\bf l})}
{d\omega^\prime d{\bf n}}=\frac{(\omega^\prime)^2}{4\pi^2c}\left|({\bf
  A}_{\rm cl}(\omega^\prime)\cdot {\bf l}^*)\right|^2.
\end{equation}
Here the vector amplitude of emission, ${\bf
  A}_{\rm cl}(\omega^\prime)$, is given by the following equation:
$$
{\bf A}_{\rm cl}(\omega^\prime,{\bf n})=\frac{e}c\int_{-\infty}^{+\infty}{{\bf
    v}(t)\exp\left(i\omega^\prime\left\{t-\frac{[{\bf n}\cdot{\bf x}(t)]}c\right\}\right)dt},
$$
see Eq.(14.67) in Ref.\cite{jack}. We express $d{\cal E}_{\rm
  rad}/(d\omega^\prime d{\bf n})$ in terms of the time integral of the radiation loss rate, $dI_{\rm cl}/(d\omega^\prime
d{\bf n})$, which is related to the unit of time, the element of a solid
angle, and the frequency interval, and is summed over polarizations:
$$
\frac\partial{\partial t}\left[\sum_{\bf l}
{\frac{d{\cal E}_{\rm rad} }
{d\omega^\prime d{\bf n} } 
}\right]=
\frac{dI_{\rm cl}(t)}
{d\omega^\prime d{\bf n} }.
$$
The spectral and angular distribution of the radiation loss rate
is given by the Fourier integral:
$$
\frac{dI_{\rm cl}(\tau)}{d\omega^\prime
d{\bf n}}=-
\frac{e^2(\omega^\prime)^2}
{4\pi^2c{\cal E}(\tau)}
\int_{-\infty}^{+\infty}
\left[p\left(\tau+\frac\zeta2\right)\cdot p\left(\tau-\frac\zeta2\right) \right]
\times
$$
$$
\times
\exp\left\{
ic
\int_{\tau-\zeta/2}^{\tau+\zeta/2}
{[k^\prime\cdot p(\tau^\prime)]d\tau^\prime}\right\}d\zeta.
$$
The cogent feature of the particle relativistic motion in strong
laser fields is that the emitted radiation is abruptly beamed about the 
direction of the velocity vector, ${\bf p}(\tau)/|{\bf p}(\tau)|$. 
Therefore, the angular spectrum of
emission can be approximated with the Dirac function:
$$
\frac{dI_{\rm cl}(\tau)}{d\omega^\prime
d{\bf n}}=\delta^2\left({\bf
  n}-\frac{\bf p}{|{\bf p}|}\right)\frac{dI_{\rm
    cl}(\tau)}{d\omega^\prime}.
$$
In the frequency spectrum of
emission,
$$
\frac{dI_{\rm cl}(\tau)}{d\omega^\prime}=
\frac{e^2\omega^\prime}
{2\pi c{\cal E}^2(\tau)}
\int_{-\infty}^{+\infty}
\frac1\zeta\left[p\left(\tau+\frac\zeta2\right)\cdot p\left(\tau-\frac\zeta2\right) \right]
\times
$$
$$
\times
\sin\left[
\frac{\omega^\prime}{{\cal E}(\tau)}
\left(\frac\zeta2+
\int_{\tau-\zeta/2}^{\tau+\zeta/2}
\left\{[ p(\tau)\cdot p(\tau^\prime) ]-1 \right\} d\tau^\prime
\right)
\right]
d\zeta,
$$ 
for relativistically strong wave
field, satisfying
Eq.(\ref{eq:strong}), the sine function oscillates fast, so that the main contribution to the integral determining 
the emission spectrum comes
from a brief time interval with small values of $\zeta$, resulting in
a {\it universal} emission spectrum: 
\begin{equation}\label{spectrum}
\frac{dI_{\rm cl}(\tau)}{d\omega^\prime}=
I_{\rm cl}\frac{Q_{\rm cl}(r)}{\omega_c},\qquad I_{\rm cl}=-\frac{2e^2(f_{Le}
\cdot f_{Le})}{3m_e^2c^3},
\end{equation} 
\begin{equation}\label{Qr0}
Q_{\rm cl}(r)=
\frac{9\sqrt{3}}{8\pi}r
\int_{r}^\infty{K_{5/3}(r^\prime)dr^\prime},\qquad
r=\frac{\omega^\prime}{\omega_c}, \label{eq:classicspectrum}
\end{equation}
\begin{equation}\label{omegac}
\omega_c=
{\cal E}\chi.
\end{equation}
Here $Q_{\rm cl}(r)$ is the 
unity-normalized 
spectrum of the
gyrosynchrotron emission, such that $\int{Q_{\rm cl}(r)dr}=1$ and
$K_\nu(r)$ are  
MacDonald functions. 
We use the dimensionless photon frequency, $\tilde{\omega}^\prime=\hbar
\omega^\prime/(m_ec^2)$, 
the characteristic frequency, $\tilde{\omega}_c=\hbar \omega_c/(m_ec^2)$,
and the dimensionless wave vector, $\tilde{\bf k}^\prime=\hbar{\bf
  k}^\prime$,
for emitted $\gamma$-photons and omit tildes in the formulae. Both the
radiation energy loss rate, $I_{\rm cl}$, and the QED-strength
parameter,
\begin{equation}\label{eq:chifirst}
\chi= \frac32\frac{\lambdabar_C\sqrt{-(f_{Le}\cdot f_{Le})}}{m_ec^2},
\end{equation} 
are expressed in terms of the 4-square of the Lorentz 4-force:
$f^\mu_{Le}={\cal E}({\bf f}_{Le}\cdot{\bf u}/c,{\bf
  f}_{Le})$, where
${\bf u}=c{\bf p}/\sqrt{1+{\bf p}^2}$ is the velocity and 
$
{\bf f}_{Le}=e{\bf E}+\frac{e}{c}[{\bf u}\times{\bf B}]
$ is the Lorentz three-force.  

Thus, the acceleration of
electrons by the laser pulse must be accompanied by 
gyrosynchrotron-like 
emission spectrum 
(which 
is actually observed -
see Refs.\cite{Rousse,kneip}). The general character of such  emission
spectrum had been noted in Ref.\cite{Rousse} (this
comment may be also found in Sec. 77 in Ref.\cite{ll}). 
The material of the present subsection was published 
in Ref.\cite{PRE}. 
\subsubsection{Equation for the radiation emission and transport}
The above considerations 
justify the method for
calculating the high frequency emission as described in Ref.\cite{our}
(see also \cite{Rousse}). In addition to
calculating the electromagnetic fields on the grid using 
a Particle-In-Cell (PIC)   
scheme, one can also account for the higher-frequency (subgrid)
emission, by calculating its  
instantaneously
radiated spectrum. Compared
with direct calculation of the RHS of 
Eq.(\ref{eq:Jackson}) (the means of calculating the emission used, for example, 
in Ref.\cite{spie},\cite{t0}) the 
approach
suggested here, 
though mathematically equivalent, may be 
decidedly more efficient.
 
Generally,
the Radiation Transport Equation (RTE, cf. Ref.\cite{ZR}) should be
solved for the radiation energy density, related to the element of
volume, $dV$ (herewith the symbol $\sum_{\bf l}$ is omitted): 
$$
{\cal I}=\frac{d{\cal E}_{\rm rad}}{d\omega^\prime d{\bf n}dV}.
$$
An electron located at the point, ${\bf x}={\bf x}_e(t)$,
contributes to the right hand side (RHS) of the RTE as follows:
$$
\frac{\partial {\cal I}}{\partial t}+c({\bf
  n}\cdot\nabla){\cal I}=\sum_e{\frac{I_{\rm cl}Q_{\rm cl}(r)}{\omega_c}
\delta^2\left({\bf n}-\frac{\bf
  p}p\right)\delta^3({\bf x}-{\bf x}_e)}.
$$
The LHS of the RTE describes the radiation transport, while in the 
RHS, in addition to the emission source,
there should be the terms accounting for the 
radiation absorption and scattering. However, at $\chi\ll1$ 
and at realistic plasma densities these effects may be neglected. 
Under these circumstances the RTE can be easily integrated over time
and space, giving:
\begin{equation}\label{eq:NonMod}
\frac{d{\cal E}_{\rm rad}}{d\omega^\prime d{\bf n}}=\int_0^t{\left(\sum_e{\frac{I_{\rm cl}Q_{\rm cl}(r)}{\omega_c}
\delta^2({\bf n}-\frac{\bf
  p}p)}\right)dt}
\end{equation}
Since Eq.(\ref{eq:NonMod}) depends 
on frequency only via $Q_{\rm cl}(r)$, one can calculate instead of
Eq.(\ref{eq:NonMod}) an integral as follows 
$$
\frac{d{\cal E}^{(m)}_{\rm rad}}{d{\bf n} d\bar{\omega}}=\int_0^t{\left[\sum_e{\frac{I_{\rm cl}}{\omega_c}\delta^2\left({\bf n}-
\frac{{\bf p}}{p}\right)
\delta\left(\log\bar{\omega} -\log\omega_c\right)}\right]d t}.
$$  
Once this modified spectrum has been integrated over the whole
simulation time, a true spectral distribution can be recovered using a simple
convolution as follows:
\begin{equation}\label{convolution}
\frac{d{\cal E}_{\rm rad}(\omega^\prime,{\bf n})}{d{\bf n} d\omega^\prime}=
\int{ Q_{\rm cl}\left(\frac{\omega^\prime}{\bar{\omega}}\right)
\frac{d{\cal E}^{(m)}_{\rm rad}(\bar{\omega},{\bf n})}
{d{\bf n} d\bar{\omega}}d\log\bar{\omega}}.
\end{equation}
Manifestly,
the result is the same, 
which allows one to avoid calculating the spectrum,
$Q_{\rm cl}(r)$, at each time step. 
\subsubsection{Equation for electron motion: accounting the
  radiation back-reaction}
Here we use the equation of motion for a radiating electron as derived in
Refs.\cite{our,mine}. In 4-vector form this equation may be written
for 
the electron 4-momentum, $p^\alpha$, normalized per $m_ec$, in terms of the 
Lorentz 4-force, $f^\alpha_{Le}=eF^{\alpha\beta}p_\beta$ and the field 4-tensor, 
$F^{\alpha\beta}=({\bf E},{\bf B})$:
\begin{equation}\label{eq:pwithj}
\frac{dp^\alpha}{d\tau}=\frac{e}{m_ec^2}F^{\alpha\beta}\frac{dx_\beta}{d\tau}-
\frac{I_{\rm cl}p^\alpha}{mc^2},
\end{equation}
\begin{equation}\label{eq:dxdtau}
\frac{dx^\alpha}{d\tau}=cp^\alpha+\frac{\tau_0f^\alpha_{Le}}{m_e}, \qquad \tau_0=\frac{2e^2}{3m_ec^3}.
\end{equation} 
Three-vector formulation of Eqs.(\ref{eq:pwithj}-\ref{eq:dxdtau}) is:
\begin{equation}\label{eq:3plus1}
\frac{d{\bf p}}{dt}=\frac{{\bf f}_{Le}}{m_ec}+\frac{e[\bar{\bf u}\times{\bf B}]}{m_ec^2}-
\frac{{\bf u}{\cal E}^2(\bar{\bf u}\cdot {\bf f}_{Le})}{m_ec^3},
\end{equation}  
$$
\frac{d{\bf x}}{dt}={\bf u}+\bar{\bf u}, \quad
\bar{\bf u}=\frac{\tau_0}{m_e}\frac{{\bf f}_{Le}-{\bf u}({\bf u}\cdot {\bf f}_{Le})/c^2}{1+\tau_0({\bf u}\cdot {\bf f}_{Le})/(mc^2)}.
$$
\subsubsection{Comparison with the Landau-Lifshitz equation.}
Many authors simulate motion of an emitting electron using the equation
as suggested by Landau and Lifshitz (LL - see Eq.(76.3) in
\cite{ll}),  
motivating a comparison
with the approach we use. To simplify
the formulae, we introduce the 4-velocity, $u^i$, and normalize the field tensor: 
$$u^\alpha=\frac{1}c\frac{dx^\alpha}{d\tau},\qquad
\tilde{F}^{\alpha\beta}=\frac{\tau_0eF^{ik}}{m_ec}=\frac{2\alpha F^{ik}}{3E_S}.$$ 
All 4-vector equations in this paragraph are written
without indices and the tensor multiplication is denoted with dot-product
and/or powers of tensor, e.g. $\tilde{F}\cdot u=\tilde{F}^{ik}u_k$, 
$\tilde{F}^2\cdot u=\tilde{F}\cdot\tilde{F}\cdot
u=\tilde{F}^{ik}\tilde{F}_{kl}u^l$ etc. Now
we re-write the LL equation:
\begin{equation}\label{eq:ll}
\frac{du}{d\tau}=\frac1{\tau_0}\tilde{F}\cdot\left(u+\tilde{F}\cdot
u\right)+\frac{d\tilde{F}}{d\tau}\cdot
  u-\frac{u}{\tau_0}(u\cdot\tilde{F}^2\cdot u).
\end{equation}
and compare it with Eqs.(\ref{eq:pwithj},\ref{eq:dxdtau}):
\begin{equation}\label{eq:our1}
\frac{dp}{d\tau}= \frac1{\tau_0}\tilde{F}u-
\frac{p}{\tau_0}(p\cdot\tilde{F}^2\cdot p),\qquad
u=p+\tilde{F}\cdot p.
\end{equation}
Solving the momentum from the second of Eqs.(\ref{eq:our1}), 
$
p=\sum_{n=0}^\infty{
(-\tilde{F})^n\cdot u}.
$ Accounting for the anti-symmetry of the field tensor,
the first of Eqs.(\ref{eq:our1}) may be re-written for
4-velocity and 4-acceleration, similarly to Eq.(\ref{eq:ll}):
\begin{eqnarray}
\frac{du}{d\tau}&=&\frac1{\tau_0}\tilde{F}\cdot\left(u+\tilde{F}\cdot
u\right)+\nonumber\\
&+&\frac{d\tilde{F}}{d\tau}\cdot\sum_{n=0}^\infty{(-\tilde{F})^n}\cdot
  u-\frac{u}{\tau_0}(u\cdot\sum_{n=1}^\infty{\tilde{F}^{2n}}\cdot u).\label{eq:lllike}
\end{eqnarray}
The only  
distinction from Eq.(\ref{eq:ll}) is that in
Eq.(\ref{eq:lllike}) the infinite series are present, while in
Eq.(\ref{eq:ll}) there are only starting terms 
of these series. 

How large is the difference {\it numerically}?
For the second series one may evaluate both the total sum:
$$
u\cdot\sum_{n=1}^\infty{\tilde{F}^{2n}}\cdot
u=p\cdot\tilde{F}^2\cdot
p=\left(\frac23\alpha\right)^2
\left(\frac49\chi^2\right),
$$ 
and the residual sum, omitted in Eq.(\ref{eq:ll}):
$$
u\cdot\sum_{n=2}^\infty{\tilde{F}^{2n}}\cdot
u
=\left(\frac23\alpha\right)^4
\left(
\frac49\chi^2\frac{E^2-B^2}{E_s^2}+\frac{({\bf
E}\cdot{\bf B})^2  }{E_s^4}\right).
$$
The residual sum is reduced by a factor, $(2\alpha/3)^2(E^2-B^2)/E_S^2$, which is
small according to Ineq.(\ref{eq:smallinv}).

How {\it theoretically} important is the distinction between the two
approaches? We discussed this issue in Refs.\cite{our,mine} and
noted that the LL equation conserves
neither the generalized momentum of electron nor the total
energy-momentum of the system consisting of an emitting electron, the
external field and the radiation. Another distinction is that
the LL 
approach maintains the identity, $u^2=1$, while 
Eqs.(\ref{eq:our1}) maintain
more important identity, $p^2=1$, 
turning
to the Dirac equation in the limit of QED-strong field. For 
the
square of the 4-velocity we obtain:
\begin{equation}\label{eq:u2}
u^2=p^2-p\cdot\tilde{F}^2\cdot p\approx 1-1.05\cdot10^{-5}\chi^2,
\end{equation}
which is not exactly 
unity, 
but in
QED-weak fields, $\chi\ll1$, the distinction is negligible.

The {\it computational} advantage of Eqs.(\ref{eq:our1}) as compared to
the LL equation, is, first, the higher efficiency: compare the compact
expression for three-force in Eq.(\ref{eq:3plus1}) with that given in
 \cite{ll} (see section 77, problem 2). Second, the numerical scheme
for Eq.(\ref{eq:3plus1}) is more reliable, as it is bound  
to yield 
total energy conservation. 
Thus, even for fields in the QED weak regime, the use of 
Eqs.(\ref{eq:our1}) is more suitable than the use of the LL equation.
\subsection{
Analytical solution}
\label{sec:analytical}  
Pertaining to the validation against a semi-analytical 
theory, we begin
by describing the spectrum of emission 
from an electron in the field of 
a 
1D circularly polarized wave. 
A
constant wave amplitude, $a_0$  
is assumed to be {\it below} the radiation-dominated
regime. In this case 
${\bf p}_\perp\approx-{\bf a}$, so that  
${\cal E}^2-p_\|^2=1+a_0^2$. The {\it modified} spectrum can be expressed
in terms of the characteristic frequency, which is 
a
function of the
current value of the electron energy:
\begin{equation}\label{omegacw}
\frac{\omega_c}{\omega_{c0}}=1+\frac{({\cal E}-p_\|)^2}{1+a_0^2}\ ,
\end{equation}
as well as the parameter,
$\omega_{c0}$ which is introduced as the following function of the wave
amplitude and frequency:
\begin{equation}
\omega_{c0}=\frac34\omega_{\rm 2Ry}\alpha^2
\left(a_0+a_0^3\right).
\end{equation}
The maximum frequency of emission is determined by the
initial momentum of electron: 
$$
\frac{\omega_{c{\rm max}}}{\omega_{c0}}=1+\frac{({\cal
    E}-p_\|)^2_{\xi=0}}{1+a_0^2}\, .
$$
Then, $\xi^*(\xi)$ is a normalized phase:
\begin{equation}\label{xistar}
\xi^*=\left(\frac23\alpha^3\omega_{\rm 2Ry} a_0^2\sqrt{1+a_0^2}\right)\xi,  
\end{equation}
For the whole pulse the total normalized phase,  
$$\xi^*_\infty=\frac23\alpha^3 a_0^2\sqrt{1+a_0^2}\xi_{\rm max}\approx\frac23
\left(\frac{J}{J_p}\right)^{3/2}\frac{\xi_{\rm max}}{(\omega_{\rm 2Ry})^{2}},
$$ 
characterizes the capability of the pulse to arrest the
counterpropagating electron by means of the radiation back-reaction.
Particularly, a pulse of duration corresponding to $\xi^*_\infty\ge1$
arrests an electron of any energy.
The modified spectrum has a shape close to a power-law (see
derivation details in 
\cite{ourAIP}):
\begin{equation}\label{analytical}
\frac{d{\cal E}^{(m)}_{\rm rad}}{d\bar{\omega}} 
=\frac{m_ec^2\sqrt{1+{\bf
      a}^2}}{4\omega_{c0}}\frac{\bar{\omega}/\omega_{c0}}{\left(\bar{\omega}/\omega_{c0}-1\right)^{3/2}},
\end{equation}
\begin{equation}
\frac{\omega_{c{\rm min}}}{\omega_{c0}}\le\frac{\bar{\omega}}{\omega_{c0}}\le
\frac{\omega_{c{\rm max}}}{\omega_{c0}},
\end{equation}
where $\omega_{c{\rm min}}$ should be found from Eq.(\ref{omegacw}), for given $\xi^*_\infty$:
\begin{equation}
\frac{\omega_{c{\rm min}}}
{\omega_{c0}}=
1+
\left(\frac{1}{\sqrt{\omega_{c{\rm max}}/\omega_{c0}-1}}+\xi^*_\infty\right)^{-2}.
\end{equation}
The true (transfromed) spectrum can be obtained from the modified
spectrum as in Eq.(\ref{analytical}) by applying a convolution 
transformation following Eq.(\ref{convolution}). The longer the pulse, 
the more softened and broadened the
radiation spectrum becomes (see Fig.1).

\begin{figure}
\includegraphics[scale=0.5]{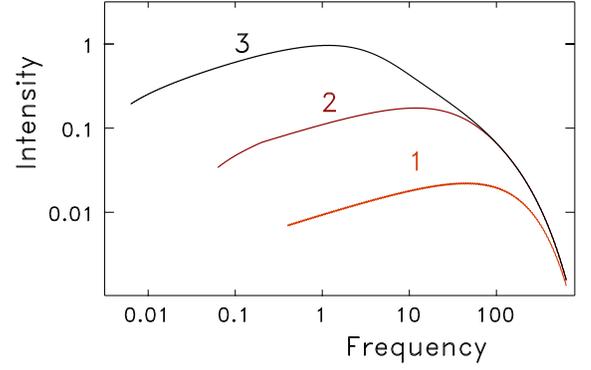}
\caption{The shape of normalized spectra, 
$(d{\cal E}_{\rm rad}/d\omega^\prime)\cdot
[4\omega_{c0}/(m_ec^2\sqrt{1+{\bf a}^2})]$, 
versus the normalized frequency,
$\omega^\prime/\omega_{c0}$, 
for different pulse duration. 
The figure is scalable, particular choice of 
physical parameters, may be the following: $|{\bf a}|=50$, 
${\cal E}=180$ MeV, pulse durations are:
5 fs (curve 1), 36 fs (curve 2) and 220 fs (curve 3). 
The spectrum broadening and softenning
is due to the radiation reaction. In the absence of this reaction  
curve 1 without changing its shape 
would scale proportionally to the pulse duration. 
A zero value of $\log \omega^\prime$ corresponds to $\approx 150$ keV.
}
\label{fig_1_ELI}
\end{figure}
\subsection{Numerical model}
Now we introduce the following normalized variables: 
$$
\tilde{t}=\omega t,\qquad \tilde{{\bf x}}=\omega{\bf x}/c, \qquad \tilde{\bf u}=\frac{\bf
  u}c,
$$
$$
\tilde{\bf E}=\frac{|e|{\bf E}}{m_ec\omega},\qquad \tilde{\bf
  B}=\frac{|e|{\bf B}}{m_ec\omega},\qquad\tilde{\bf j}
=\frac{4\pi|e|{\bf j}}{m_e c\omega^2}.
$$
Note that the electric current density, $\tilde{\bf j}$ is normalized
per $|e|n_{\rm cr}c$, where  $n_{\rm cr}=m_e\omega^2/(4\pi e^2)$ is the
critical density. 
Below, we use these dimensionless variables and omit tildes in
notations. 
The equations of motion for electrons and positrons read:
$$
\frac{d{\bf p}_{e,p}}{dt}=
{\bf f}_{Le,p}\mp[\bar{\bf u}_{e,p}\times{\bf B}]-{\bf
        u}_{e,p}{\cal E}_{e,p}^2\left({\bf f}_{Le,p}\cdot\bar{\bf u}_{e,p}\right),
$$
$$
\frac{d{\bf x}_{e,p}}{dt}={\bf u}_{e,p}+\bar{\bf u}_{e,p},\quad{\cal E}_{e,p}=\sqrt{1+{\bf p}^2_{e,p}},\quad {\bf u}_{e,p}=\frac{{\bf p}_{e,p}}{\cal E}_{e,p},
$$
the normalized Lorentz force and $\bar{\bf u}$ being:
$$
{\bf f}_{Le,p}=\mp\left({\bf E}+[{\bf u}_{e,p}\times{\bf B}]\right),\qquad\varepsilon=\tau_0\omega=\frac23\alpha^3\omega_{\rm 2Ry}
$$
\begin{equation}\label{calcubar}
\bar{\bf u}_{e,p}=\varepsilon\frac{{\bf f}_{Le,p}-{\bf
    u}_{e,p}({\bf u}_{e,p}\cdot{\bf f}_{Le,p})}{1+\varepsilon({\bf
    u}_{e,p}\cdot{\bf f}_{Le,p})}.
\end{equation}
For reference we also provide the energy equation:
\begin{equation}\label{epenergy}
\frac{d{\cal E}_{e,p}}{dt}=\mp
\left(\left[{\bf u}_{e,p}+\bar{\bf
  u}_{e,p}\right]\cdot{\bf E}\right)
-{\cal E}_{e,p}^2\left({\bf f}_{Le,p}\cdot\bar{\bf u}_{e,p}\right),
\end{equation}
For ions with the charge number, $Z$, and
the mass, $M_i$, the momentum is normalized per $M_ic$, so that in their
equation of motion the electron-to-ion mass ratio comes:
\begin{equation}
\frac{d{\bf p}_i}{dt}=\frac{Z}{M_i/m_e}\left({\bf E}+[{\bf u}_i\times{\bf B}]\right),
\end{equation}
\begin{equation}
\frac{d{\bf x}_i}{dt}={\bf u}_i, \qquad{\bf u}_i=
\frac{{\bf p}_i}{\sqrt{1+{\bf p}^2_i}}.
\end{equation}
Below we assume that $Z=1$ and $M_i=M_p$ is the proton mass. 
The normalized Maxwell equations read:
\begin{equation}\label{eq:Maxwell}
\frac{\partial{\bf E}}{\partial t}+ {\bf j}=\nabla\times{\bf B},\qquad \frac{\partial{\bf B}}{\partial t}=-\nabla\times{\bf E}.
\end{equation}
\subsubsection{Macroparticles and their current}
 We assume that the rectangular grid splits the 
 computational domain into the control volumes 
 (cells), $\Delta V=\prod{\Delta x}_k$. If the electron density 
 equals $n_{\rm cr}$, there are 
$n_{\rm cr}\Delta V$ electrons per cell. 
The latter number is typically 
very large, so that the plasma electrons cannot be simulated
individually and they are combined into macroparticles with a large 
number of ``electrons-per-particle'', $N_{epp}$. In 
a plasma of 
critical density the number of (macro)particles per cell is: 
$N_{ppc}=n_{\rm cr}\Delta V/N_{epp}$.

The electron current density inside the given cell is expressed in
terms of the sum over electron macroparticles in this cell. As long as
the electric current density is normalized per $n_{\rm cr}c$, the 
contribution to the latter sum from each macroparticle equals
$-(1/N_{ppc})d{\bf x}_e/dt$. On adding the contributions
from positrons and protons, we obtain:
\begin{equation}\label{current}
{\bf j}=
\frac{-\sum_e{({\bf u}_e+\bar{\bf u}_e)}+
 \sum_p{({\bf u}_p+\bar{\bf u}_p)}+
 \sum_i{{\bf u}_i}}{N_{ppc}}.
\end{equation}
\subsubsection{Energy integral and energy balance.}
We now establish the relationship between the energy integral 
and the finite sum which represents this integrals in simulations. Particularly, 
the field energy may be calculated as the total of
point-wise field magnitudes squared:
$
{\cal E}_{\rm field}=\frac12\sum_{cells}{({\bf E}^2+{\bf B}^2)},
$ 
which is by a factor of ${\cal E}_0=m_ec^2(n_{\rm cr}\Delta V)$ different
from the dimensional field energy. 
Now consider the total plasma energy, which  
includes the particle energy
as well:
$$
{\cal E}_{\rm plasma}=
{\cal E}_{\rm
  field}+\frac{\sum_{cells}{\left(\frac{M_i}{m_e}\sum_i{\cal E}_i+\sum_e{\cal E}_e+\sum_p{\cal E}_p\right)}}{N_{ppc}}.
$$ 
Eqs.(\ref{epenergy},\ref{eq:Maxwell},\ref{current}) give: 
$
d({\cal E}_{\rm plasma}+{\cal E}_{\rm rad})/dt=0,
$
where
\begin{equation}
\frac{d{\cal E}_{\rm rad}}{dt}=
\frac{\sum_{cells}{\sum_{e,p}{{\cal
          E}^2_{e,p}\left({\bf f}_{Le,p}\cdot\bar{\bf u}_{e,p}\right)}}}{N_{ppc}}
\end{equation}
is the radiation energy loss rate. Therefore, the contribution from
electrons and positrons to the dimensionless radiation energy at each
time step, $\Delta t$, is calculated as ${\cal
          E}^2_{e,p}\left({\bf f}_{Le,p}\cdot\bar{\bf
  u}_{e,p}\right)\Delta t/N_{ppc}$. 
Once integrated over the simulation time,
the radiation 
energy may be converted to physical units on multplying
it by a factor of ${\cal E}_0$. 
\subsubsection{Algorithmic implementation}
The algorithmic changes
to the standard PIC scheme are
minimal as long as we ignore the radiation transport and only integrate
over time the energy emitted by electrons (and positrons, if any). To
collect the radiation, we introduce the
energy bins (an array) ${\cal E}_{{\rm rad}\,ijk}(\log(\bar{\omega})_i,\theta_j,\varphi_k)$,
which discretize the {\it modified} frequency-angular spectrum of
emission. Inside the desired interval of the photon energies we
introduce the logarithmic grid, $\log(\bar{\omega})_i$, equally spaced
with the step, $\Delta\log(\bar{\omega})$. We also introduce  
a grid,
$\theta_j,\varphi_k$, for 
the 
two polar angles of 
the 
spherical coordinate
system, with $\Delta{\bf n}_{jk}$ being the element of solid
angle: $\Delta{\bf n}_{jk}=\sin(\theta_j)\Delta\theta\Delta\varphi$. 

To calculate both the spectrum of emission and the radiation 
back-reaction we modify only that part of the PIC algorithm which
accounts for the electron motion. Specifically, we employ the standard
leapfrog numerical scheme which involves, among 
others, the
following stages: (1) for each electron macroparticle, 
update momentum 
through
the time step by adding the Lorentz force, following the Boris scheme:
${\bf p}^{n+1/2}_e={\bf p}^{n-1/2}_e+\Delta t\ {\bf f}_{Le}({\bf
  E}^{n},{\bf B}^{n})$; (2) solve the energy and the velocity from
the updated momentum: 
${\cal E}^{n+1/2}_e=
\sqrt{1+(
{\bf p}^{n+1/2})^2_e}$, 
${\bf u}^{n+1/2}_e={\bf p}^{n+1/2}_e/{\cal E}^{{n+1/2}}_e$; (3) use the calculated velocity
to update the particle coordinates: 
${\bf x}^
{n+1}_e=
{\bf x}^{n}_e+
{\bf u}^{n+1/2}_e\Delta t$ and account for the 
contribution of the
electric current 
element, $-{\bf u}^{n+1/2}_e\Delta t$, to the Maxwell 
equations. Again, these stages are
standard and may be found in \cite{birdsall}. We introduced new steps
into this algorithm between stages (2) and (3) as follows.

\begin{itemize}
\item[2.1]
Once stage (2) is done, recover the Lorentz force: 
${\bf f}_{Le}=(
{\bf p}^{n+1/2}-{\bf p}^{n-1/2})/\Delta t$.
\item[2.2]
Find $\chi=\frac32
\omega_{\rm 2Ry}
\alpha^2{\cal E}^{n+1/2}
\sqrt{{\bf f}^2_{Le}-({\bf f}_{Le}\cdot{\bf u}^{n+1/2})^2}$.
\item[2.3]
Find $\bar{\bf u}_e$ by putting ${\bf f}_{Le}$ and ${\bf
  u}_{e}^{n+1/2}$ into Eq.(\ref{calcubar}). 
\item[2.4]
Calculate $\omega_c={\cal E}^{n+1/2}_e\chi$ and find the discrete value
of $\log(\bar{\omega})_i$ most close to $\log{\omega_c}$. Find the
angles, $\theta_j,\varphi_k$ 
closest to the direction  of ${\bf p}^{n+1/2}$. 
Add the radiation energy into the proper bin
$$
{\cal E}_{{\rm rad}\,ijk}\rightarrow{\cal E}_{{\rm rad}\,ijk}+
\frac{({\cal E}^{n+1/2})^2(
{\bf f}_{Le}\cdot\bar{\bf u}_e)\Delta
    t}{\omega_cN_{ppc}\Delta\log(\bar{\omega})\Delta{\bf n}_{jk}}.$$
\item[2.5]
Add the radiation force:
${\bf p}_{e}^{n+1/2}\rightarrow {\bf p}_{e}^{n+1/2}+\Delta t\left\{-[\bar{\bf
    u}_e\times{\bf B}^{n}]-{\bf u}_{e}^{n+1/2}({\cal E}^{n+1/2}_{e})^2(
{\bf f}_{L{e}}\cdot\bar{\bf u}_{e})\right\}$.
\item[2.6]
Find
$
{\bf u}^{n+1/2}_e=
{\bf p}_e^{n+1/2}/\sqrt{1+({\bf p}_e^{n+1/2})^2}+\bar{\bf u}_e
$
and use this velocity through stage (3).  
\end{itemize}
Note that the algorithm modification is applied
only to electrons (positrons), keeping unchanged the ion motion as well
as the fields.

The frequency-angular spectrum may be reduced to the frequency 
one: ${\cal E}_{{\rm rad}\,i}=\sum_{jk}{{\cal E}_{{\rm rad}\,ijk}\Delta{\bf n}_{jk}}$, to the 
angular one: 
${\cal E}_{{\rm rad}\,jk}=\sum_{i}{{\cal E}_{{\rm
    rad}\,ijk}\bar{\omega}_i\Delta\log(\bar{\omega})}$ or to the total
radiation energy: ${\cal E}_{\rm rad}=\sum_{ijk}{{\cal E}_{{\rm
    rad}\,ijk}\bar{\omega}_i\Delta\log(\bar{\omega})}\Delta{\bf
  n}_{jk}$. 

While postprocessing the results, we apply the convolution
transformation, (\ref{convolution}), to the radiation spectrum and
multiply it by ${\cal E}_0$.  
The resulting 
spectrum, $d{\cal E}_{\rm rad}/(d{\bf n}d\omega^\prime)$, is a function 
of $\log[\hbar\omega^\prime/(m_ec^2)]$.   
\begin{figure}
\includegraphics[scale=0.4]{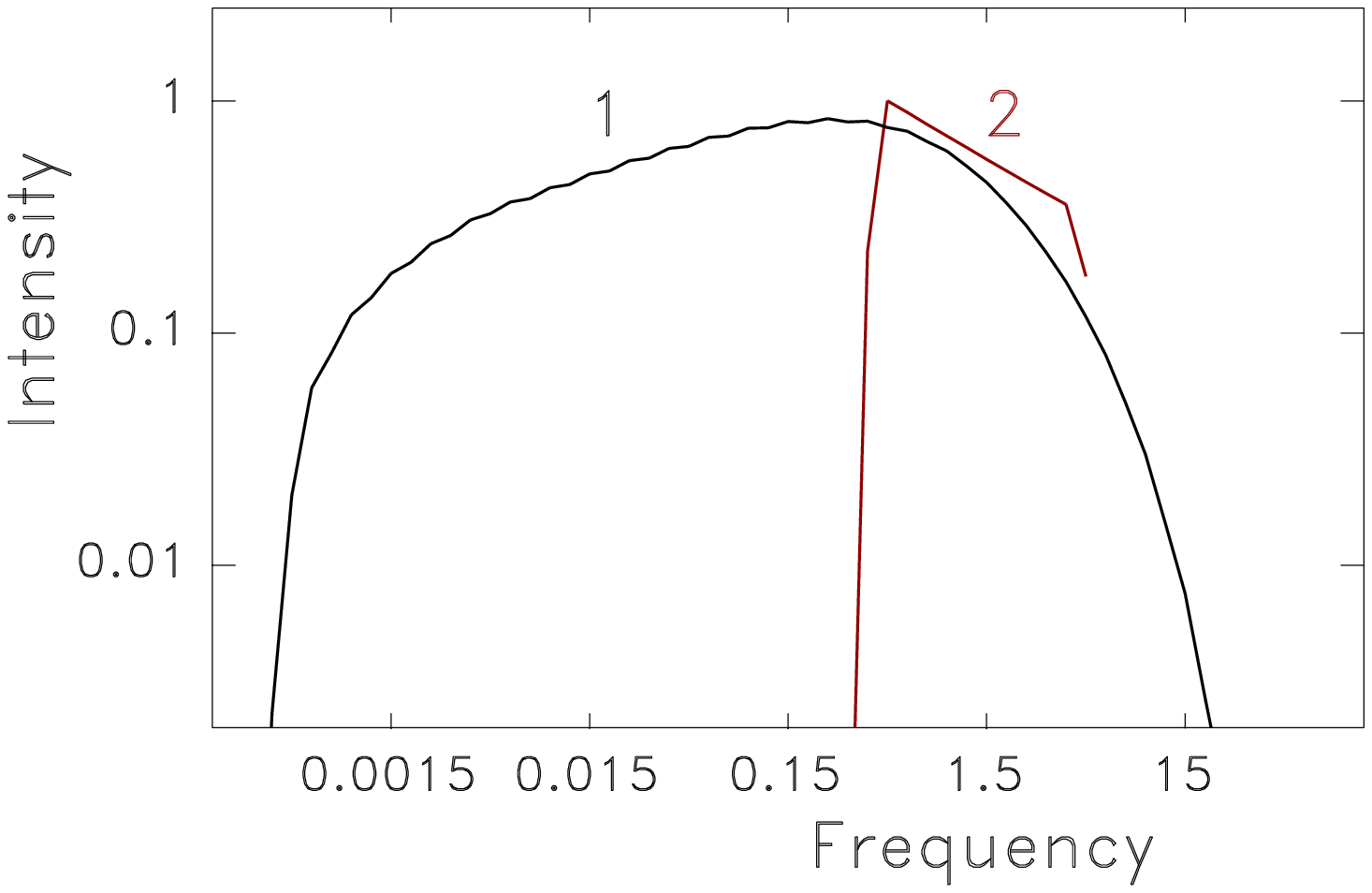}
\includegraphics[scale=0.3]{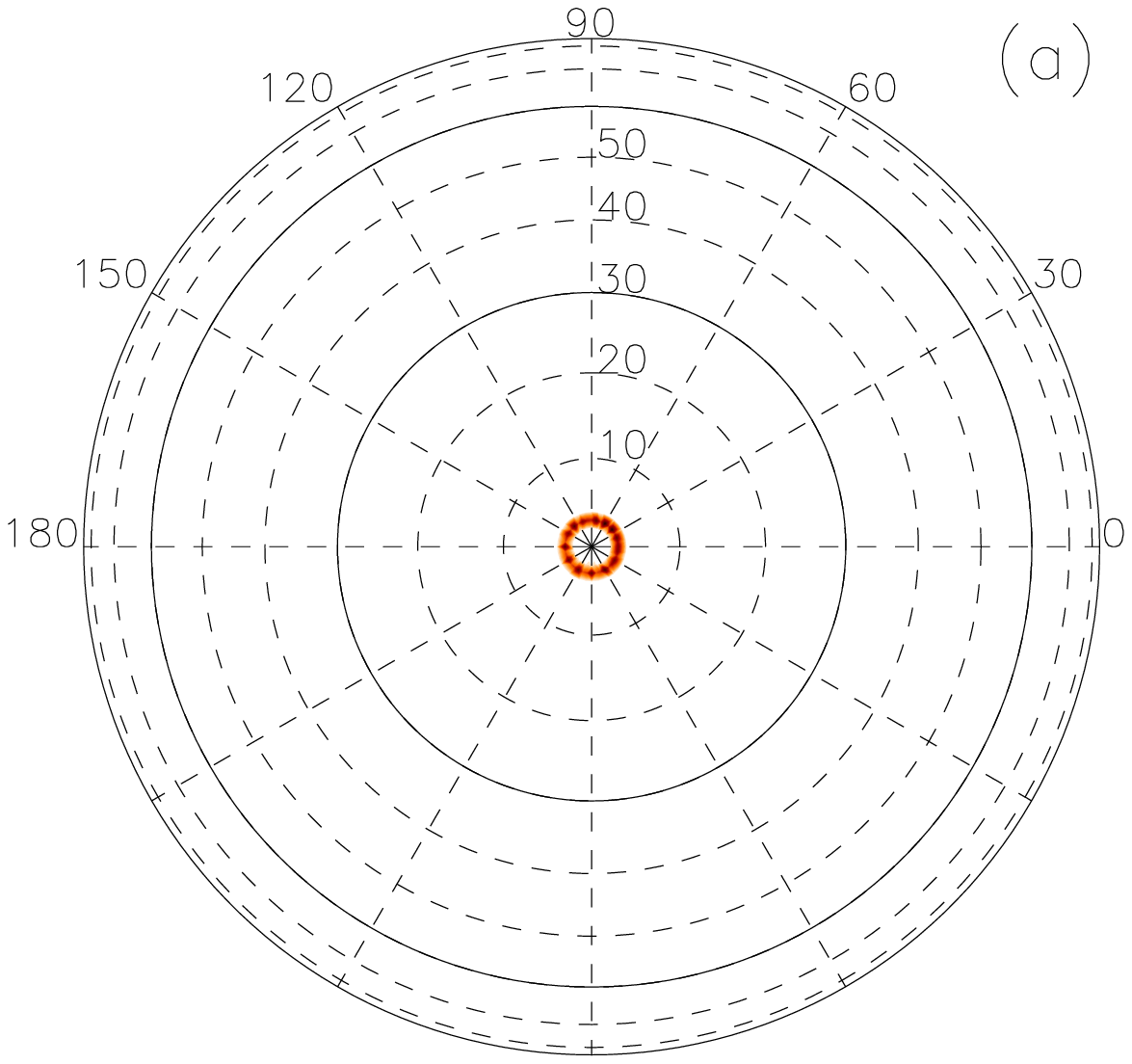}
\includegraphics[scale=0.3]{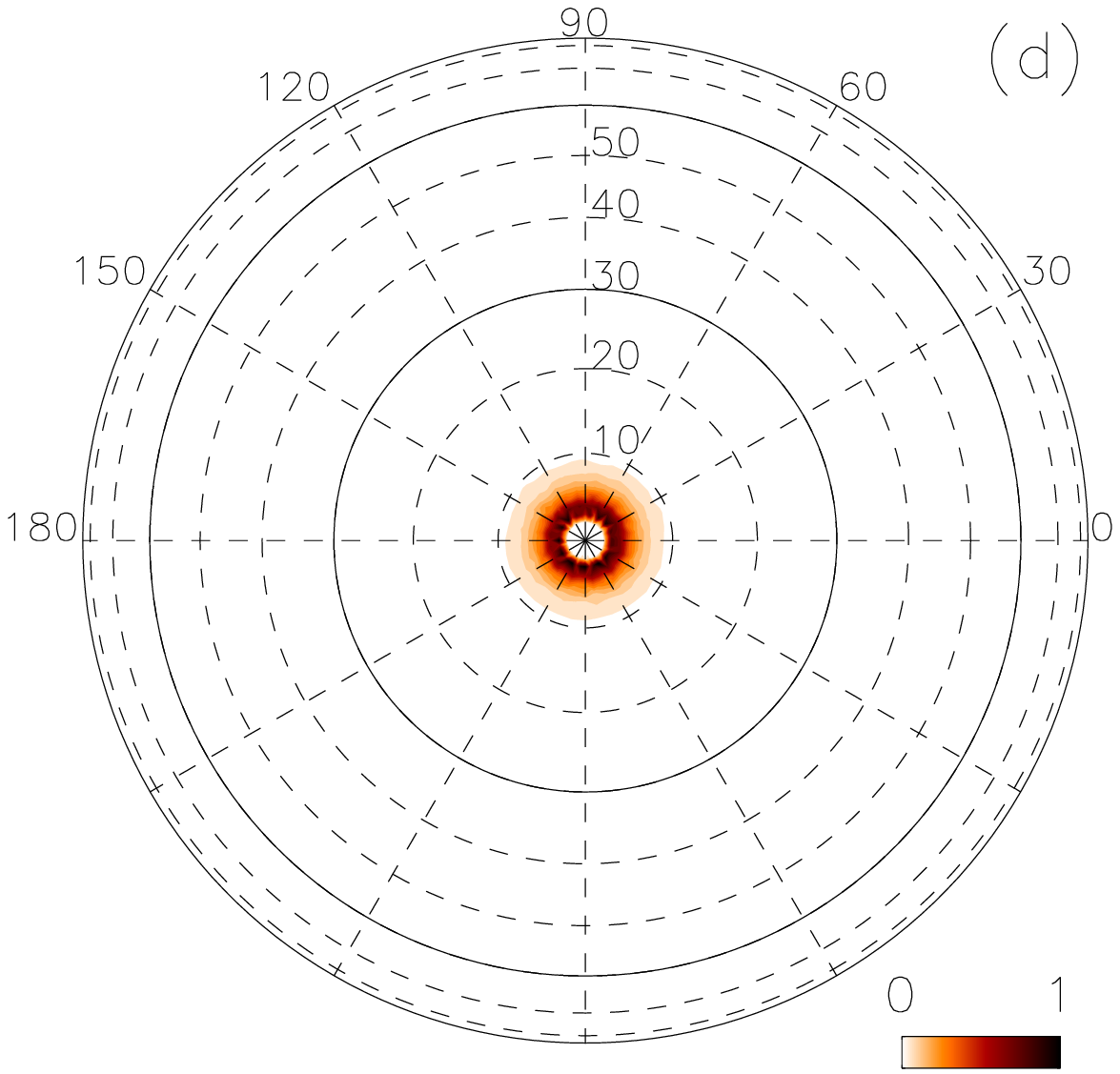}
\caption{Test simulation result: (a): radiation energy spectrum (line 1), $d{\cal E}_{\rm rad}/d\omega^\prime$, 
and  the  modified spectrum, $d{\cal E}_{\rm rad}/d\bar{\omega}$ 
(line 2); 
(b),(c): the angular distribution of the radiation at instants: (b)
t=1T, (c) t=100T, where $T=2\pi/\omega$. }
\label{fig_3_ELI}
\end{figure}
\subsection{Simulation result}
The analytical solution presented in Sec.\ref{sec:analytical}
has been used to benchmark the numerical scheme. In the test simulation
electrons with an initial momentum, 
$p_\|=300$, propagate toward the laser pulse with 
sharp (2$\lambda$) fronts. The circularly polarized laser pulse has amplitude, 
$|{\bf a}|=15$, and duration, $100(2\pi/\omega)$. Interacting with the pulse, 
the particles radiate energy, finally approaching momentum of
$p_\|\approx130$.

In Fig.\ref{fig_3_ELI}(a)
the spectrum of the resulting radiation 
$d{\cal E}_{\rm rad}/d\omega^\prime$ is shown. We also provide the modified spectrum
(the distribution over $\bar{\omega}=\omega_c$), which is close to 
satisfying a
power law, 
and in a full agreement with the analytical solution. 
In Fig.\ref{fig_3_ELI}(b)-(c) 
typical evolution of 
the
angular radiation distribution, $d{\cal E}_{\rm
  rad}/d{\bf n}$, 
is provided for the same simulation. 
One can see that the 
majority  of the radiation is concentrated 
in a narrow angle with respect to the direction of backscattered light (0$^o$). 
A softer 
part of the
radiation 
exhibits a wider angular distribution and becomes less intense 
[Fig. \ref{fig_3_ELI}(c)]. 
\section{QED-moderate fields ($\chi\sim 1$) } 
When $\chi$ is not small ($J\sim 3\cdot 10^3{\rm W/cm}^2$), QED
effects come into play. Here we describe how to {\it extend}
the methods used above towards finite $\chi$. This is achieved by 
applying realistic QED spectra of emission as derived in
Ref.\cite{PRE}, with the radiation force modified accordingly. 

This approach is applicable as long as we
ignore
the onset of some {\it new} effects which are only pertinent to
QED-strong fields. Specifically, while employing the radiation force,
$dp^\mu_{\rm rad}/d\tau$, it is admitted that the change in the 
electron momentum, $d\tau\cdot dp^\mu_{\rm rad}/d\tau$, within the 
infinitesimal time interval, $d\tau$, is 
also infinitesimal.
This 
'Newton's law' approximation is pertinent to classical physics and
it ignores the point that the change in the electron momentum at 
$\chi\sim1$ is essentially finite because of the finite momentum of
emitted photon. The approximation, however, is highly efficient and allows 
one
to avoid time-consuming statistical simulations. Its error
tends to zero as $\chi\rightarrow0$, and it is sufficiently small at 
$\chi\sim1$. 

Another effect which we ignore in this section is the pair production
due to $\gamma$-photon absorption in the strong laser field. This
neglect allows us to avoid solving the computationally intense
radiation transport problem.
\subsection{Theoretical notes}                         
\subsubsection{Emission spectrum}
The emission probability found in Ref.\cite{PRE} within the framework of QED
can be reformulated in
a form similar to Eq.(\ref{eq:Jackson}). The polarized
part of emission may be reduced to Eq.(\ref{eq:Jackson}) with the
modified 
vector amplitude: 
$$
{\bf A}_{\rm QED}(\omega^\prime)=\sqrt{
\frac{1}{C_{fi}}} 
{\bf A}_{\rm cl}\left(
\frac{\omega^\prime}{C_{fi}}\right),
\qquad
C_{fi}=\frac{(k\cdot p_i)}{(k\cdot p_f)},
$$
subscript $i$ and $f$ denoting  
the parameters of 
 an 
 electron prior to and
after the emission of a single photon, and: 
\begin{equation}\label{eq:pol}
\frac{dI_{\rm QED}^{\rm pol}(\omega^\prime)}
{d\omega^\prime}=C_{fi}\frac{dI_{\rm cl}(\frac{\omega^\prime}{C_{fi}})}
{d\omega^\prime}=\frac{C_{fi}I_{\rm cl}}{\omega_c}Q_{\rm cl}
\left(\frac{r}{C_{fi}}\right).
\end{equation}
Within the framework of QED the electron  
possesses not only an
electric charge, but also a magnetic moment associated with its spin. 
Usually the spin is assumed to be depolarized (as is done in Ref.\cite{PRE}), 
and, accordingly, 
a depolarized
contribution to the emission 
appears:
\begin{equation}\label{eq:depol}
\frac{dI^{\rm depol}_{\rm
  QED}}{d\omega^\prime}=\frac{I_{\rm
    cl}(\tau)}{\omega_c}
\frac{9\sqrt{3}}{8\pi}\left(1-C_{fi}\right)^2
\frac{r}{C_{fi}}K_{2/3}\left(\frac{r}{C_{fi}}\right).
\end{equation}
Thus, the QED effect in the emission from 
an
electron in a strong
electromagnetic field
reduces to 
a downshift in frequency accompanied by 
an extra
contribution from the magnetic moment of 
electron. The universal emission spectrum in QED-strong fields 
is given by
the total of Eqs.(\ref{eq:pol},\ref{eq:depol}):
$$
\frac{dI_{\rm QED}}{d\omega^\prime}=\frac{I_{\rm
    cl}}{\omega_c}q(\chi)Q_{\rm QED}(r,\chi), \quad I_{\rm QED}=I_{\rm
  cl}q(\chi),
$$
where $Q_{\rm QED}=Q^\prime_{\rm QED}/q$ is the normalized by unity
spectrum, $q(\chi)=\int_0^\infty{Q^\prime_{\rm QED}(r,\chi)dr}\le1$ is the 
normalization parameter, the spectrum before normalization is:  
$$
Q^\prime_{\rm QED}(r,\chi)=
\frac{9\sqrt{3}}{8\pi}r\left[
\int\limits_{r_\chi}^\infty{K_{5/3}(r^\prime)dr^\prime}+\chi^2rr_\chi K_{2/3}(r_\chi)\right],
$$
and $r_\chi=r/(1-\chi r)$. 
\begin{figure}
\includegraphics[scale=0.38]{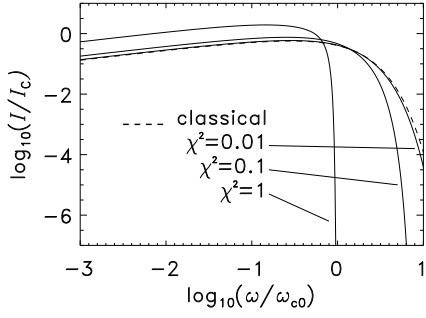}
\caption{Emission spectra for various values of $\chi$.
}
\label{fig_2}
\end{figure}
\begin{figure}
\includegraphics[scale=0.38]{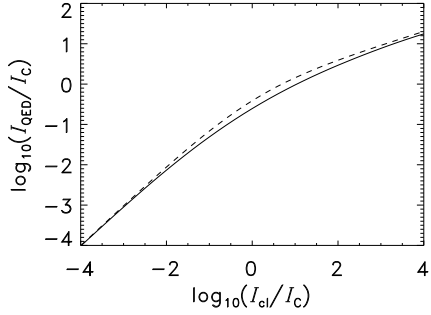}
\caption{Emitted radiation power in the QED approach {\it vs} classical 
(solid); an interpolation formula $I_{\rm QED}=I_{\rm
    cl}/(1+1.04\sqrt{I_{\rm cl}/I_C})^{4/3}$ (dashed). Here,
  $I_C=I_{\rm cl}/\chi^2$.}
\label{fig_3}
\end{figure}
\subsubsection{Equation for electron motion: accounting the
  radiation back-reaction.}
As long as the QED effects modify emission,
\begin{equation}\label{QEDmod}
Q_{\rm cl}(r)\rightarrow Q_{\rm QED}(r,\chi),\qquad I_{\rm
  cl}\rightarrow I_{\rm QED},
\end{equation}
the radiation back-reaction needs to be revised accordingly. 
In Refs.\cite{mine,our} it was noted,  that QED is not compatible 
with the traditional approach to the radiation force in classical
electrodynamics, while
Eqs.(\ref{eq:pwithj},\ref{eq:dxdtau}) may be employed at 
finite value of $\chi$ on substituting $I_{\rm QED}$ for $I_{\rm
  cl}$. Alternatively, 
within the framework of QED the radiation back-reaction may be found by
integrating the 4-momentum carried away with the emitted photons 
and that absorbed from the external electromagnetic
field in the course of emission. For a 1D wave field 
this procedure gives the following equation (see Ref.\cite{PRE}):
\begin{equation}\label{eq:our}
\frac{dp^\alpha}{d\tau}=\frac{f^\alpha_{Le}}{m_ec}+\frac{I_{\rm QED}}{m_ec^2}\left[\frac{k^\alpha}{(k\cdot p)}-p^\alpha\right],
\end{equation}
and in such field $F^{\alpha\beta}F_{\beta\mu}p^\mu/(p_\nu
  F^{\nu\beta}F_{\beta\mu}p^\mu)=k^\alpha/(k\cdot p)$, $k^\alpha$ being
  the wave 4-vector. 
Eq.(\ref{eq:our}) coincides with Eqs.(\ref{eq:pwithj},\ref{eq:dxdtau}), in which
the substitution (\ref{QEDmod}) is made, or, the same, the
three-vector formulation
for $\bar{\bf u}$ is applicable to the 1D wave field, if the substitution is done as follows: 
\begin{equation}\label{tauQED}
\tau_0\rightarrow\tau_0\frac{I_{\rm QED}}{I_{\rm
    cl}}=\tau_0q(\chi).
\end{equation} 
Although this approach is derived for the 1D field, we may apply it to
an arbitrary 3D focused field. An argument in favor of this
generalization is that the property of a 1D wave, $(k\cdot
k)=0$, which is used while deriving Eq.(\ref{eq:our}), holds as an
approximation for any field. Indeed, on calculating the 4-square of
$F^{\alpha\beta}F_{\beta\mu}p^\mu/(p_\nu
  F^{\nu\beta}F_{\beta\mu}p^\mu)$,
which 4-square is similar to
$(k\cdot k)/(k\cdot p)^2$, we find:
$$
\frac{p\cdot F^4\cdot p}{(p\cdot
  F^2\cdot p)^2}=
\frac94\chi^{-2}\frac{E^2-B^2 }{E_S^2}+
\frac{81}{16}\chi^{-4}\frac{({\bf E}\cdot{\bf B})^2
  }{E_S^4}\ll1,
$$ 
the inequality holds at $\chi\ge1$ according to
Ineq.(\ref{eq:smallinv}). 

Another criterion
which should be checked at $\chi\ge1$ is the requirement for the
difference, $(1-u^2)\propto\chi^2$, as in Eq.(\ref{eq:u2}) to be
small. Applying the substitution (\ref{tauQED}) to (\ref{eq:u2}) we
find that $(1-u^2)\le 2\cdot 10^{-6}$, reaching its maximal value at 
$\chi\sim3.4$. This 'error' is negligible, even if one assumes, than any
theory allowing $u^2\ne1$, is erroneous.    
\subsection{Analytical result}
\begin{figure}
\includegraphics[scale=0.4]{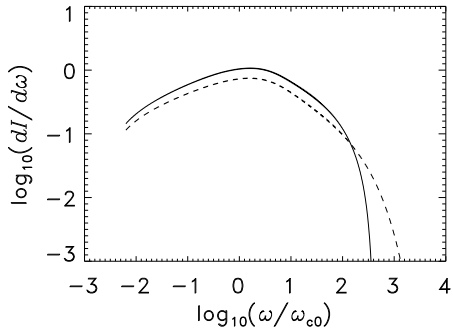}
\caption{The emission spectrum for 600 MeV electrons interacting with 
30-fs laser pulses of intensity $2\cdot 10^{22} W/cm^2 $: with (solid) or 
without (dashed) accounting for the QED effects. Here $\hbar\omega_{c0}\approx 1.1$ MeV for $\lambda=0.8\mu$m.
}
\label{fig_5}
\end{figure}

In Fig.\ref{fig_5} we show the emission spectrum for an electron interacting with a
laser pulse (see \cite{PRE} for detail). We see that the QED
effects essentially modify the spectrum even 
with laser intensities which are already achieved.
\subsection{Numerical model.
}
As long as the QED spectrum of emission depends on $\chi$,  the bins
for collecting the radiation energy
should be refined: 
${\cal E}_{{\rm rad}\,ijkl}(\bar{\omega}_i,{\bf n}_{jk},\chi_l)$. Once
for a given electron (or positron) the parameter $\chi$ is calculated; 
the discrete value of $\chi_l$ should be found most close to $\chi$. 
Then, parameter $\varepsilon$ should be found following 
Eq.(\ref{tauQED}), $\varepsilon=q_l\tau_0\omega$, using 
pre-tabulated value, $q_l=q(\chi_l)$. Then, $\bar{\bf u}_{e}$ should be
expressed in terms of $\varepsilon$ and the radiated energy should be
added to a proper energy bin:
$$
{\cal E}_{{\rm rad}\,ijkl}\rightarrow{\cal E}_{{\rm rad}\,ijkl}+
\frac{({\cal E}^{(n+1/2)}_e)^2(
{\bf f}_{Le}\cdot\bar{\bf u}_e)\Delta
    t}{\omega_cN_{ppc}\Delta\log(\bar{\omega})\Delta{\bf n}_{jk}}.$$ 
While postprocessing the results, a convolution similar to
(\ref{convolution}) should be applied with $\chi_l$-dependent spectra:
$$
\frac{d{\cal E}_{\rm rad}(\omega^\prime,{\bf n})}{d{\bf n}
  d\omega^\prime}
={\cal E}_0
\sum_{\chi_l}{
\int{ Q_{\rm QED}\left(\frac{\omega^\prime}{\bar{\omega}},\chi_l\right)
{\cal E}_{{\rm rad}\,ijkl}
d\log\bar{\omega}}}.
$$
\subsection{Simulation results}
\begin{figure}
\includegraphics[scale=0.4]{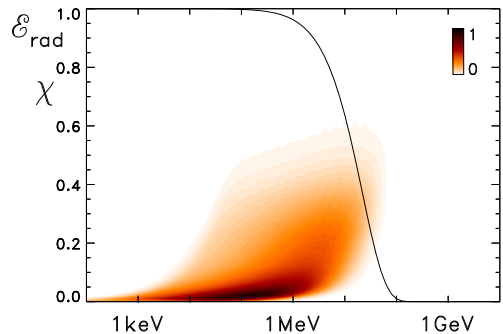}
\caption{Backscattered light in simulation of the interaction of
  laser pulse of intensity $2\times 10^{23}$W/cm$^2$ with plasma of
  density $6.5\times 10^{22}$cm$^{-3}$.  
For each $\chi_l$ (vertical axis) the
convolution integral (the term in the above formula for convolution) 
is calculated and presented as a function of
$\hbar\omega^\prime$ (horizontal axis).
The line shows the total emitted energy 
as a function of the cutoff photon energy (i.e. the integral spectrum). 
}
\label{fig_6}
\end{figure}
In a 1D simulation presented in Fig.\ref{fig_6} a linearly polarized laser 
pulse with a step-like profile having 2-$\lambda$ front and amplitude
$a=300$ interacts with plasma of density $n_0=30n_c$ during
50 cycles. About half of laser energy is converted to high energy
photons. The data for backscattered photons indicate that values of
$\chi\approx 1$ are achieved. These values are reasonable, as 
the energy of electrons moving toward
the pulse is as high as 180MeV, and the vector potential approach
values $a=400$ due to superposition of incident and reflected light. 
One can see that $\sim$65\% of emitted photons exceed 10MeV, 
and 96\% are above 1MeV. 
\section{QED-strong fields ($\chi\gg1$)}
\subsection{Theoretical notes}
In cases
where  $\chi\gg1$ one needs to solve the RTE in order to account
for 
$\gamma$-photon absorption in strong fields. This 
may be done using the Monte-Carlo 
method, in which the radiation field is 
evaluated statistically
(see Refs.\cite{nay09,mas09,bae09}). Instead of the radiation energy 
density the photon distribution function may be introduced as follows:
\begin{equation}
f_\gamma({\bf x},\omega^\prime,{\bf n})=\frac{{\cal I}({\bf
    x},\omega^\prime, {\bf n})}{m_ec^2\omega^\prime}.
\end{equation}
Similar to the way the electron macroparticles represent the
electron distribution 
function, the photon marcoparticles may
be employed to simulate the photon distribution function. To simulate
emission, the photons are created with their momentum selected
statistically. The photon propagation in the direction of ${\bf n}$ is 
simulated in the same way as for electrons. The absorption with the known probability  
is also simulated statistically.  

Now we may split the radiation for the part which may be
treated in the way we followed so far (see Sec.II-III) and for the
Monte-Carlo photons. We choose a parameter,
$\chi^*\sim(0.05-1)$ and assume that: (1) an electon with 
$\chi\le\chi^*$ contributes only to $d{\cal E}_{\rm
  rad}/(d\omega^\prime d{\bf n})$; while (2) for an electron 
with $\chi>\chi^*$ the {\it regular} spectrum of emission,
$Q^\prime_{\rm QED}(r,\chi)$ (which is normally truncated at
$r=1/\chi$), is now truncated at $r=\chi^*/\chi^2<1/\chi$ and 
the emission of photons with $\chi^*/\chi^2\le r\le\chi$, or, the
same,
$
\frac{\chi^*}{\chi}\le\frac{\omega^\prime}{{\cal E}}\le1,
$
is treated statistically. The regular radiation loss rate
as well as the contribution to the radiation force $\propto-{\bf
  p}I_{\rm QED}$ should be both reduced by a factor of 
$q_t(\chi)/q(\chi)$ at $\chi>\chi^*$, where the truncated spectrum integral equals:
$
q_t(\chi)=\int_0^{\chi^*/\chi^2}{Q^\prime_{\rm QED}(r,\chi)dr}
$. The normalized by unity 
trucnated spectrum is:
$Q_t(r,\chi)=\frac{q(\chi)Q_{\rm QED}(r,\chi)}{q_t(\chi)},\,r<\chi^*/\chi^2$. The
emission probability to be used at $\chi>\chi^*$, $r>\chi^*/\chi^2$ may
be found
in \cite{PRLsuppl}:  
$$
dW_{e\rightarrow e,\gamma}=
\frac{
\kappa K_{2/3}(r_\chi)
+\int\limits_{r_\chi}^\infty{K_{5/3}(r^\prime)dr^\prime}}{\pi\sqrt{3}\alpha\omega_{\rm 2Ry}}
d\left(\frac{\omega^\prime}{
{\cal E}}\right)\frac{d(\omega t)}{{\cal E}},
$$
where $r_\chi=\frac{\omega^\prime/{\cal E}}{\chi(1-\omega^\prime/{\cal
    E})}$, $\kappa=\frac{(\omega^\prime/{\cal E})^2}{1-\omega^\prime/{\cal
    E}}$ and for a 1D wave field we use an equation, $\frac{d(\omega
  t)}{\sqrt{3}\alpha\omega_{\rm
    Ry}}=\frac{\sqrt{3}\alpha d\xi}{2\chi}\left|\frac{d{\bf
    a}}{d\xi}\right|$. If the  emission probability is averaged over time or
over an ensemble, we return to the above spectrum of emission:
$\omega^\prime m_ec^2\left<dW_{e\rightarrow e,\gamma}/(drdt)\right>=
I_{\rm QED}Q_{\rm QED}(r,\chi)$, Here we apply  
the formula, $\frac{1}{\sqrt{3}\alpha\omega_{\rm
    Ry}}=\frac{9\sqrt{3}I_{\rm QED}}{8\chi^2q(\chi)\omega m_ec^2}$,
which is also used in the numerical scheme. 
\subsection{Semi-analytical solution}  
In \cite{ourPRL} we demonstrated that as long as the distribution 
functions, $f_{e,p,\gamma}$, for electrons, positrons and photons 
in a 1D wave field are integrated over the transversal components 
of momentum, their evolution is described by simple kinetic 
equations with the collision integrals. 
We solved these equation numerically. This choice of initial conditions 
corresponds to the 46.6 GeV SLAC electron beam and 
the laser intensity of $J\approx 5\cdot 10^{22}\,{\rm W/cm}^2$
for $\lambda=0.8\mu$m,
to be achieved soon. As long as the Monte-Carlo method 
is not used,
the numerical results, such as the total pair production, may be used
to benchmark the numerical scheme described here.   
\subsection{ Numerical model}
The modification of the numerical scheme as used in Sec.III is needed
only for electrons and positrons with $\chi>\chi^*$. The radiation 
energy added 
to the proper energy bin is
corrected as follows:
$$
{\cal E}_{{\rm rad}\,ijkl}\rightarrow{\cal E}_{{\rm
    rad}\,ijkl}+\frac{q_t(\chi_l)}
{q(\chi_l)}
\frac{({\cal E}^{(n+1/2)}_e)^2(
{\bf f}_{Le}\cdot\bar{\bf u}_e)\Delta
    t}{\omega_cN_{ppc}\Delta\log(\bar{\omega})\Delta{\bf n}_{jk}},$$
and the same  
correction factor is applied to the second term in braces
at algorithm stage 2.5. After stage 2.5 a probable hard photon emission from the electron with $\chi>\chi^*$ is
accounted, using the probability:
$$
\frac{dW_{e\rightarrow e,\gamma}}{d(\omega^\prime/{\cal E})}=\delta
w\left(\frac{
\omega^\prime}{\cal E},\chi\right),\quad\delta=\frac{({\cal E}^{(n+1/2)}_e)^2(
{\bf f}_{Le}\cdot\bar{\bf u}_e)\Delta
    t}{\chi^2{\cal E}^{(n+1/2)}},
$$ 
$$
w\left(\frac{
\omega^\prime}{\cal E},\chi\right)=
\frac{9\sqrt{3}}{8\pi q(\chi)}\left[\kappa K_{2/3}(r_\chi)
+\int\limits_{r_\chi}^\infty{K_{5/3}(r^\prime)dr^\prime}\right],
$$
The total probability of emission is
given by a complete integral:
$
W_{e\rightarrow e,\gamma}=
\delta
\int_{\chi^*/\chi}^1{
w\left(\frac{\omega^\prime}{\cal E},\chi\right)d\left(\frac{\omega^\prime}{\cal E}\right)}$. 
Both within the QED perturbation theory and within the Monte-Carlo
scheme $W_{e\rightarrow e,\gamma}$ is assumed to be less than one. 
The probability of no emission 
equals $1-W_{e\rightarrow e,\gamma}\ge0$. The partial probability, $W_{e\rightarrow e,\gamma}
(\omega^\prime<\omega^\prime_0)$, for the emission with the photon
energy 
not exceeding the given value,
$\omega^\prime_0$, is given by the incomplete probability integral:
$$
W_{e\rightarrow e,\gamma}
(\omega^\prime<\omega^\prime_0)=
\delta
\int_{\chi^*/\chi}^{\omega^\prime_0/{\cal E}}{w\left(\frac{
\omega^\prime}{\cal E},\chi\right)d\left(\frac{\omega^\prime}{\cal E}\right)}
$$
Therefore, for given $\delta$ and $\chi$ and for a randomly generated number, 
$0\le {\rm rnd}<1$ one can solve
$\omega^\prime/{\cal E}$ from an integral equation as follows (see
detail in Appendix B): 
\begin{equation}\label{eq:event}
\int^{
\omega^\prime/{\cal E}}_{\chi^*/\chi}{
w\left(z,\chi\right)dz
}=\frac{\rm rnd}\delta\le \int^{1}_{\chi^*/\chi}{
w\left(z,\chi\right)dz
},
\end{equation}
if the gambled value of ${\rm rnd}$ does not exceed
$W_{e\rightarrow e,\gamma}$: $0\le{\rm rnd}\le W_{e\rightarrow
  e,\gamma}$. Otherwise 
 (if $W_{e\rightarrow e,\gamma}<{\rm
  rnd}\le1$) 
the extra emission
does not occur. With calculated $\omega^\prime/{\cal E}$, the
emission is accounted for by creating a new photon macroparticle with
the
momentum, ${\bf p}^{(n+1/2)}_e(\omega^\prime/{\cal E})$ and an electron
recoil is accounted for by reducing the electron momentum, 
${\bf p}^{(n+1/2)}_e\rightarrow{\bf p}^{(n+1/2)}_e(1-(\omega^\prime/{\cal E}))$.
   
\subsubsection{Photon propagation and absorption} 
The new element of
the numerical scheme is the photon macroparticle, which simulates
$(n_{\rm cr}\Delta V)/N_{ppc}$ real photons. Its propagation with 
dimensionless velocity equal to ${\bf n}$ is treated in the same way as
for electrons and ions. 

If the photon escapes the computational domain, its energy should be
accounted for while calculating the total emission
from plasma. For this purpose we introduce the
energy bins, ${\cal E}^{(tot)}_{{\rm rad}\,ijk}(\log({\omega^\prime})_i,\theta_j,\varphi_k)$,
such that the logarithmic equally spaced grid for the photon energy,
$\log({\omega}^\prime)_i$, and the polar angle grid coincide with those
introduced above. The contribution from the escaping photon with 
total energy, $\omega^\prime m_ec^2n_{\rm cr}\Delta V/N_{ppc}$, should
be added to the bin with the closest
$\log({\omega^\prime})_i,\theta_j,\varphi_k$, with the macroparticle
energy being converted to the spectral energy density by dividing by $\Delta
\omega^\prime =\omega^\prime\Delta \log({\omega^\prime})$:
$$
{\cal E}^{(tot)}_{{\rm rad}\,ijk}\rightarrow {\cal E}^{(tot)}_{{\rm
    rad}\,ijk}+
\frac{m_ec^2(n_{\rm cr}\Delta V)}{N_{ppc}\Delta \log({\omega^\prime})}
$$

The photon absorption with 
electron-positron-pair creation is
gambled in the same way as the emission (see details in Appendix B). 
Other absorption 
mechanisms may be also included. In postprocessing
the simulation results, the softer
$\gamma$-photon emission should be added to the total radiation spectrum:
$$
\frac{d{\cal E}_{\rm rad}(\omega^\prime,{\bf n})}{d{\bf n}
  d\omega^\prime}
={\cal E}^{(tot)}_{{\rm
    rad}\,ijk}+m_ec^2(n_{\rm cr}\Delta V)\times
$$
$$
\times\sum_{\chi_l}{
\int{ Q_{t}\left(\frac{\omega^\prime}{\bar{\omega}},\chi_l\right)
{\cal E}_{{\rm rad}\,ijkl}
d\log\bar{\omega}}}
$$
\subsection{Simulation result}
Repeating the test simulation as described in Sec.III.D and applying
the Monte-Carlo scheme at $\chi>\chi^*=0.1$, and without photon absorption,
we notice only an increase in the fluctuations of the high-energy portion of the radiation spectrum. In this region, the photons are statistically 
underrepresented, the number of particles 
per $\Delta \log(\omega^\prime)$ being small.
\section{Conclusion}
Thus, the range of 
field intensities which may be simulated with 
good accuracy using
the described tools is now extended towards the 
intensities as high as $(2-3)\cdot10^{23}{\rm W/cm}^2$. In such fields,
which are typical for the radiation-dominated regime of the 
laser-plasma 
interaction, the suggested scheme is validated against 
a semi-analytical solution. Different versions of the equation of the
emitting particle motion are compared and their proximity is 
demonstrated. 

Extension of the model for
QED-strong fields can be easily incorporated  
into the scheme. 
The emission spectra are substantially modified by 
QED effects and
simulation results for 
a realistic laser-plasma interaction 
are provided. 

For the QED-strong field regime of laser-plasma interaction the
Monte-Carlo method should be used to simulate emission-absorption of
harder $\gamma$-photons. Although such simulations are doable (see
\cite{ner11}), more work on the scheme validation is needed.   
\section*{Appendix A. MacDonald functions}
The MacDonald functions allow solutions for the following integrals:
$$
\int\limits_{-\infty}^{+\infty}{\cos\left[\frac32r\left(\frac{z^3}3+z\right)\right]dz}=\frac2{\sqrt{3}}K_{1/3}(r),
$$
(see Eq.(8.433) in \cite{gr})
$$
\int\limits_{-\infty}^{+\infty}{z\sin\left[\frac32r\left(\frac{z^3}3+z\right)\right]dz}=\frac2{\sqrt{3}}K_{2/3}(r),
$$
$$
\int\limits_{-\infty}^{+\infty}{\frac1z\sin\left[\frac32r\left(\frac{z^3}3+z\right)\right]dz}=-\frac2{\sqrt{3}}
\int\limits_r^{+\infty}{K_{1/3}(r^\prime)dr^\prime},
$$
Using the known relationships, 
$
2\frac{dK_\nu(r)}{dr}=-K_{\nu-1}(r)-K_{\nu+1}(r), 
$
 $K_{-\nu}(r)=K_\nu(r)$, more integrals may be
reduced to the MacDonald functions, particularly:
$$
\int\limits_{-\infty}^{+\infty}{\frac{1+2z^2}z\sin\left[\frac32r\left(\frac{z^3}3+z\right)\right]dz}=\frac2{\sqrt{3}}
\int\limits_r^{+\infty}{K_{5/3}(r^\prime)dr^\prime}.
$$
The advantage of the MacDonald functions is the fast
convergence of their integral representation:
$$
K_{\nu} (r)=
\int_0^{\infty}{ \exp\left[-r \cosh(z)\right]\cosh(\nu
    z) dz }, 
$$
$$
\int_r^\infty{K_{\nu} (r^\prime)dr^\prime}=
\int_0^{\infty}{ \exp\left[-r \cosh(z)\right]\frac{\cosh(\nu
    z)}{\cosh(
    z)} dz },
$$
(see Eq.(8.432) in \cite{gr}). In numerical simulations, therefore, the
MacDonald functions may be calculated by integrating their
representations using the Simpson method,
unless the argument, $z$, is very small or very large, in which case
one can employ the series and asymptotic expansions for cylindrical
functions (see \cite{gr}).
\section*{Appendix B. Event generator for emission}
To solve Eq.(\ref{eq:event})
numerically, one needs to pre-calculate the table:
$$
W_{ml}=\int^{
(\omega^\prime/{\cal E})_m}_{\chi^*/\chi}{
w\left(z,\chi_l\right)dz
}
$$
for discrete $\chi_l$ and for discrete uniformly spaced values
of $(\omega^\prime/{\cal E})_m\le1$. For known $\chi_l>\chi^*$
the positive value of $W_{ml}$  most close to ${\rm rnd}/\delta$ determines the
value of $(\omega^\prime/{\cal E})_m$.

For the absorption probability we use an analogous formule from \cite{PRLsuppl}:
$$
dW_{\gamma\rightarrow e,p}=
\frac{
\kappa K_{2/3}(r_\chi)
-\int\limits_{r_\chi}^\infty{K_{5/3}(r^\prime)dr^\prime}}{\pi\sqrt{3}\alpha\omega_{\rm Ry}}
d\left(\frac{
{\cal E}}{\omega^\prime}\right)\frac{d(\omega t)}{\omega^\prime},
$$
where $r_\chi=\frac1{\chi_\gamma(1-{\cal
    E}/\omega^\prime){\cal E}/\omega^\prime}$, $\kappa=\chi_\gamma
r_\chi$. Here, ${\cal E}$ is the  energy of the electron 
created in the $\gamma$-photon absoption together with a positron
of energy $\omega^\prime-{\cal E}$. The $\chi$-parameter for a
photon, $\chi_\gamma$, is calculated as for an electron with an energy,
$\omega^\prime$, and a dimensionless velocity, ${\bf n}$, in terms of the local
electromagnetic field intensities.

\end{document}